\documentstyle[sprocl]{article}

\def\tib{\tilde{b}}
\def\tic{\tilde{c}}
\def\tid{\tilde{d}}
\def\tig{\tilde{g}}

\def\aw{$a_\mu^w$}
\def\bw{$b_\mu^w$}
\def\cw{$c_{\mu\nu}^w$}
\def\dw{$d_{\mu\nu}^w$}
\def\ew{$e_\mu^w$}
\def\fw{$f_\mu^w$}
\def\gw{$g_{\la\mu\nu}^w$}
\def\Hw{$H_{\mu\nu}^w$}

\def\al{\alpha}
\def\be{\beta}
\def\ze{\zeta}
\def\et{\eta}
\def\la{\lambda}
\def\om{\omega}
\def\De{\Delta}
\def\Om{\Omega}

\def\fr#1#2{{{#1} \over {#2}}}

\def\X{{\hat X}}
\def\Y{{\hat Y}}
\def\Z{{\hat Z}}

\def\bea{\begin{eqnarray}}
\def\eea{\end{eqnarray}}
\def\ket#1{|{#1}\rangle}
\def\etal{{\it et al.}}

\newcommand{\beq}{\begin{equation}}
\newcommand{\eeq}{\end{equation}}
\newcommand{\rf}[1]{(\ref{#1})}

\begin{document}

\title{CPT AND LORENTZ TESTS WITH CLOCKS IN SPACE}

\author{N.E.\ RUSSELL}

\address{Physics Department, Northern Michigan University, \\
1401 Presque Isle Avenue, Marquette, MI 49855, USA\\E-mail:
nrussell@nmu.edu}

\maketitle\abstracts{ Space-based clock-comparison experiments
can provide Planck-scale sensitivity to many parameters for
Lorentz and CPT violation that are difficult to measure on Earth.
The principal advantages are
a reduced timescale for data collection,
reduced suppression for certain effects,
and access to certain parameters not possible with Earth-based experiments.
}

\section{Introduction}
The standard model of particle physics
is covariant under rotations and boosts,
which together make up the Lorentz transformations.
Violation of Lorentz symmetry\cite{cpt98}
might occur in the context of string theory,
accompanied possibly by violation of CPT
symmetry.\cite{kps}
Observation of such violations
would provide information about noncommutative field
theories,\cite{chklo}
and would represent evidence of Planck-scale physics.
It is unlikely that they could be directly observed,
but suppressed effects might be reachable with high-sensitivity experiments.
The description of observable effects
is given by a standard-model extension that allows for
violation of both Lorentz and CPT
symmetry.\cite{ck}

Clock-comparison experiments with atoms and
ions\cite{ccexpt,lh,db,dp}
are some of the most sensitive existing tests
of Lorentz and CPT symmetry in matter.
These search for violations of rotational symmetry
by monitoring the frequency variations
of a Zeeman hyperfine transition
as the quantization axis changes direction.
The usual configuration involves
comparing the frequencies of two different co-located clocks
as the laboratory rotates with the Earth.
Sensitivity to suppressed effects from the Planck scale
can be achieved with experiments of this
type.\cite{kla}
Other sectors
of the standard-model extension
can be accessed through different experiments
involving hadrons,\cite{kexpt,ak,bexpt,ckpvi,bckp}
photons,\cite{ck,cfj}
muons,\cite{vh}
and electrons.\cite{eexpt,eexpt2}

This article aims to show that
tests of Lorentz and CPT symmetry
with Planck-scale sensitivity
can be fruitfully pursued with
clock-comparison experiments
on satellites and other spacecraft.\cite{bklr}
We consider generalities of space experiments
and discuss feasible tests for some specific
orbital and deep-space missions,
including a few approved
for the International Space Station (ISS).

Certain Zeeman hyperfine transitions
are shifted in frequency by
the presence of Lorentz and CPT
violation.\cite{kla}
For a clock operating on such a transition,
contributions to these shifts
are controlled at leading order by a few parameters
denoted in the clock reference frame as
$\tib_3^w$,
$\tic_q^w$,
$\tid_3^w$,
$\tig_d^w$,
$\tig_q^w$.
Here, the superscript $w$ is $p$ for the proton,
$n$ for the neutron, and $e$ for the electron.
These quantities are particular combinations
of the basic coefficients
\aw, \bw, \cw, \dw, \ew, \fw, \gw, \Hw\
appearing in the standard-model extension.
They are related to expectation values in the underlying fundamental theory.
For example,
\beq
\tib_3^w = b_3^w -m_w d_{30}^w
+ m_w g_{120}^w -H_{12}^w \ ,
\eeq
where $m_w$ is the mass of the particle of type $w$
and the subscripts are indices defined in
a reference frame
with the $3$ direction defined as the clock quantization axis.

\section{Reference Frames}
We first consider a clock in a laboratory on the surface of the Earth.
Then the sidereal rotation of the Earth,
with period $ 23 \, {\rm h} \, 56 \, {\rm min} \simeq 2\pi/ \Om $,
gives rise to related variations in the parameters
$\tib_3^w$,
$\tic_q^w$,
$\tid_3^w$,
$\tig_d^w$,
$\tig_q^w$.
This time dependence can be found
by considering the transformation
from the clock frame with coordinates $(0,1,2,3)$
to a nonrotating frame with coordinates $(T,X,Y,Z)$.
Ideally, the nonrotating frame should be inertial,
but for practical purposes
any frame close enough to inertial
to achieve the desired experimental sensitivity
would suffice.
Frames associated with the Earth, the Sun, the Milky Way galaxy,
or the cosmic microwave background radiation
are examples of possible choices for the nonrotating frame.
In previous literature,
the nonrelativistic conversion from
the clock frame to the nonrotating frame has been considered.
However,
the high velocities attainable
in space-based experiments
make it attractive to consider also
leading-order relativistic effects due to clock boosts.
Existing experimental bounds are unaffected
by this choice of nonrotating frame
since they ignored the translational motion of the clock.
However, an Earth-centered choice
is not appropriate for relativistic experiments because
it is inertial over a limited time scale of perhaps a few days.
Alternatively,
frames centered on the Sun, the galaxy,
or the microwave background
are approximately inertial
over thousands of years.
This choice of frame must be stated when reporting bounds,
but all of these are acceptable.

We adopt the Sun-based frame as a natural choice for experiments.
It is convenient to center the
spatial origin on the Sun
with the unit vector $\Z$ parallel to the Earth's rotation axis,
$\X$, $\Y$ in the equatorial plane,
and $\X$ directed towards the celestial vernal equinox.
Time $T$ is measured from the vernal equinox in the year 2000
using a clock located at the spatial origin.
In this inertial frame,
the Earth orbits about the Sun in a plane lying at an angle of
$\et \simeq 23^\circ$ with respect to the $XY$ plane.

It will suffice to approximate the Earth's orbit as circular
with angular frequency $\Om_\oplus$
and speed $\be_\oplus$.
In the same way,
a satellite orbit about the Earth can be approximated as circular
with angular frequency $\om_s$
and speed $\be_s$.
The angle between $\Z$ and the axis of the satellite orbit
will be denoted by $\ze$,
and the right ascension angle of the ascending node of the orbit
will be denoted by $\al$.
The oblateness of the Earth and other perturbations
cause $\al$ to precess by about 4 degrees per day.

Expressed in the Sun-based frame,
the clock boost
is $\vec V(T)=d\vec X/dT$,
where the instantaneous spatial location
$\vec X(T)$ of the clock
is determined by the trajectories of the spacecraft and the Earth.
This vector $\vec V(T)$ determines the dilation
of infinitesimal time intervals in the clock frame
relative to ones in the Sun-based frame.
Effects such as
small perturbations in $\vec V(T)$
and the gravitational potential
should be included in an
accurate relation between the two times.
However,
these corrections are irrelevant
when two clocks at essentially the same location are compared.
According to conventional relativity, the clocks then keep identical time
independent of their composition.
However,
in the presence of Lorentz and CPT violation
a signal that cannot be mimicked in conventional relativity
is generated,
because two co-located clocks involving different atomic species
typically behave differently.

Depending on the flight mode of the satellite,
the orientation of the clock quantization axis
may change relative to the Sun frame.
For this article, we focus
on a flight mode and clock configuration
where the clock quantization axis
is tangential to the circular satellite trajectory about the Earth.
So, the reference frame for the clock is chosen with
1 axis pointing towards the center of the Earth,
2 axis perpendicular to the satellite orbital plane, and
3 axis parallel to the satellite motion about the Earth.
This configuration should be possible
with planned modes for the ISS.
Other spacecraft flight modes and quantization-axis orientations
can be handled by our general methodology.
It should be noted that to gain optimal sensitivity to certain components
specific quantization-axis orientations
relative to the plane of the orbit and the angular momentum vector
would be required.

By combining the boost $\vec V(T)$ of the orbiting clock
with its rotation,
the signal in the clock frame can be converted
to the Sun-based frame.
The idea is that
components of the coefficients for Lorentz violation
in the clock frame
are to be expressed in terms of components
in the Sun-based frame.
For example,
the component $b_3^w$ becomes
\bea
b_3^w &=& b_T^w \{\be_s - \be_\oplus [\sin \Om_\oplus T
(\cos \al \sin \om_s \De T
  \nonumber \\
&&
\qquad
\qquad
+ \cos \ze \sin \al \cos \om_s \De T )
- \cos \et \cos \Om_\oplus T
  \nonumber \\
&&
\qquad
\qquad
\times (\sin \al \sin \om_s \De T
- \cos \ze \cos \al \cos \om_s \De T )
  \nonumber \\
&&
\qquad
\qquad
+ \sin \et \cos \Om_\oplus T \sin \ze \cos \om_s \De T]\}
  \nonumber \\
&&
- b_X^w (\cos \al \sin \om_s \De T
+ \cos \ze \sin \al \cos \om_s \De T )
  \nonumber \\
&&
- b_Y^w (\sin \al \sin \om_s \De T
- \cos \ze \cos \al \cos \om_s \De T )
  \nonumber \\
&&
+b_Z^w \sin \ze \cos \om_s \De T,
\label{b3}
\eea
where $\De T = T - T_0$ is the time interval measured from an
agreed reference time $T_0$.
Effects such as the Thomas precession are neglected,
since the equation holds to leading order in linear velocities.
The full result for the Sun-frame observable parameter
$\tib_3^w$
involves the expression
\rf{b3} for the component $b_3^w$
as well as expressions for other coefficients.
The other observables
$\tic_q^w$,
$\tid_3^w$,
$\tig_d^w$, and
$\tig_q^w$ are found by a similar procedure.
The expressions are lengthy,
depending on various combinations of
basic coefficients for Lorentz and CPT violation,
on trigonometric functions of various angles,
on frequency-time products,
on $\be_\oplus$, and on $\be_s$.

\section{Signal Properties}
All the spatial components of
the basic coefficients for Lorentz and CPT violation
are directly accessible with space-based experiments.
Ground-based clock-comparison experiments
seeking frequency variations as the Earth rotates
are limited by the fixed rotation axis,
which means sensitivity to certain spatial components
is not possible.
For instance,
ground-based experiments
are sensitive only to the nonrotating-frame components
$\tib_X^w$, $\tib_Y^w$ of the parameter $\tib_3^w$,
and can therefore only bound a limited subset of components of
\bw, \dw, \gw, \Hw.
All spatial components can however be accessed with an orbiting satellite.
Typically,
satellites offer different sensitivity
from Earth-based experiments since their
orbital plane is tilted with respect to the
equatorial plane.
Furthermore,
the precession of the satellite orbital plane
makes it possible in principle to access all spatial directions.

Another advantage of space-based experiments
is the relatively short orbital period,
a result of their high orbital speeds.
Since the satellite orbital period $2 \pi /\om_s$ is typically
much less than the sidereal day,
the time required to collect data
can be substantially reduced.
For example,
a clock-comparison experiment on the ISS
could be completed approximately 16 times faster
than an Earth-based one,
since the ISS orbital period is about 92 min.
This better matches clock stabilities
and reduces the run time from months to days.
This makes possible
an analysis of the leading-order relativistic effects
due to the speed
$\be_\oplus \simeq 1\times 10^{-4}$
of the Earth in the Sun-based frame.
As a result, sensitivity to many more
types of Lorentz and CPT violation can be achieved.
Existing ground-based clock-comparison experiments
might take data over months,
during which the Earth's velocity vector changes direction significantly.
In space-based experiments,
this vector could be treated
as approximately constant
if the timescale for data collection
is short enough.
This would considerably simplify the experimental analysis
since the Earth could be regarded as
an inertial frame,
thus permitting direct extraction
of leading-order relativistic effects.

Sensitivity to many types of Lorentz and CPT violation
that remain unconstrained to date
could be achieved with space-based experiments.
For example, consider
a clock-comparison experiment
sensitive to the observable $\tib_3^w$ for some $w$.
In the Sun-based frame and for each particle species $w$,
this observable involves
the basic coefficients
\bw, \dw, \gw, \Hw\
for Lorentz violation,
a total of 35 independent observable components
if allowance is made for the
effect of field redefinitions.
Whereas a traditional ground-based experiment
is sensitive to 8 of these,
the same type of experiment mounted on a space platform
would acquire sensitivity to all 35.
We note that experiments could be envisaged using
an Earth-based rotating turntable
to gain access to a wider set of parameters.
The emphasis here is on understanding and optimizing
sensitivities envisaged for
planned space missions.

For Earth-based experiments,
relativistic Lorentz and CPT terms
are suppressed by the boost factor $\be_\oplus$.
However,
space-based clock-comparison experiments
would be sensitive to first-order relativistic effects
proportional to $\be_s$.
Investigating the corresponding effects
in Earth-based experiments
would be impractical,
and in any case these would be further suppressed
by a factor of $\Om/\om_s$.
For the ISS,
$\Om/\om_s$ is about $6 \times 10^{-2}$.

A seemingly counterintuitive effect exists
among the order-$\be_s$ corrections.
In space-based experiments
a dipole shift can generate a potentially detectable signal
with frequency $2\om_s$.
This is not seen in the usual nonrelativistic analysis of
ground-based clock-comparison experiments,
where signals with the double frequency $2\Om$ occur only
in quadrupole shifts.
To gain insight into this,
consider the parameter $\tib_3^w$,
which nonrelativistically
is the third component of a vector
and so would lead only to a signal with frequency $\om_s$.
However,
$\tib_3^w$ contains the component $d_{03}$.
In a relativistic treatment incorporating first-order effects
from $\be_s$ it behaves like a two-tensor
and hence can produce a signal with frequency $2\om_s$.
As an example,
when the Earth is near the northern-summer solstice,
the expression for $\tib_3^w$
in the Sun-based frame has a double frequency term
that goes like $\cos ( 2 \om_s \De T)$
with coefficient $C_2$ containing the following spatial
components of \dw:
\bea
C_2 &\supset&
\be_s \fr m 8 [ \cos 2 \al (3+\cos 2 \ze) (d^w_{XX} - d^w_{YY})
\nonumber \\
&&
+ (1- \cos 2 \ze) (d^w_{XX} + d^w_{YY} - 2 d^w_{ZZ})
\nonumber \\
&&
- 2 \sin 2 \ze (\cos \al \, (d^w_{YZ} + d^w_{ZY})
- \sin \al \, (d^w_{ZX} + d^w_{XZ}))
\nonumber \\
&&
+ (3+\cos 2 \ze)\sin 2 \al \, (d^w_{XY} + d^w_{YX})] . %
\label{2oms}
\eea
Sensitivity to all observable spatial components
of \dw could thus be achieved by observing
the $2\om_s$ frequency.

\section{Earth-Satellite Experiments}
The ISS is of special interest since it is the planned
platform for numerous scientific experiments in the near future.
For the ISS, the relevant orbital parameters include
$\be_s \simeq 3 \times 10^{-5}$
and $\ze \simeq 52^\circ$.
Instruments planned for installation
are H masers, laser-cooled Cs and Rb clocks,
and superconducting microwave cavity
oscillators.\cite{parcs,aces,race,sumo}
Advantages for experiments on the ISS include
a microgravity environment
and reduced environmental disturbances,
which are expected to lead to gains in sensitivity
compared to existing ground-based clocks.
The analysis presented here is valid for
possible Lorentz and CPT tests
with all these instruments,
except the oscillators.\cite{km}
For the present discussion,
we assume the signal clock is compared to a co-located reference clock
insensitive to leading-order Lorentz and CPT violation.
This could be an H maser tuned to its clock transition
$\ket{1,0} \rightarrow \ket{0,0}$, for example.

\subsection{Hydrogen masers}
One option would be to use an H maser as the signal clock
as well as the reference clock.
Such an experiment would be
similar to a recent ground-based Lorentz and CPT test,
which measured the maser transition
$\ket{1,\pm 1} \rightarrow \ket{1,0}$
using a double-resonance
technique.\cite{dp}
Sensitivity would be to the parameters
$\tib_3^p$ and $\tib_3^e$
in the clock frame
and the analysis in this case has the advantage
that the atoms used are relatively simple compared with
those in other atomic-clock experiments.
In addition,
the short ISS orbital period implies
that an experimental run of about a day
would be sufficient
to obtain data roughly equivalent to four months
of data taken on Earth.
For both $w=e$ and $w=p$, all spatial components
of \bw, $m_w$\dw, $m_w$\gw, \Hw\ could be sampled
by using the orbital inclination ($\ze\neq 0$)
and by repeating the experiment
with a different value of $\al$.
We estimate that several presently unbounded components would
be tested at the level of about $10^{-27}$ GeV,
while others would be tested at about $10^{-23}$ GeV.
This is based on the assumption that previous sensitivities of
about 500 $\mu$Hz can be achieved in space.
Cleaner bounds on certain
components of $m_w$\dw, $m_w$\gw
at the level of about $10^{-23}$ GeV
could be obtained by searching for a signal
at the double frequency $2\om_s$.
Planck-scale sensitivity
to about 50 components of coefficients
for Lorentz and CPT violation
that are currently unconstrained could be tested
in this way.

\subsection{Cesium Clocks}
The reference frequency for a
laser-cooled $^{133}$Cs clock
could be the usual clock transition
$\ket{4,0} \rightarrow \ket{3,0}$,
which is insensitive to Lorentz and CPT violation.
An attractive signal transition in the present context
would be a Zeeman hyperfine transition such as
$\ket{4,4} \rightarrow \ket{4,3}$.
Since the electronic configuration of $^{133}$Cs
involves an unpaired electron,
the electron-parameter sensitivity
is similar to that of the H maser.
In the Schmidt model,
the $^{133}$Cs nucleus is a
proton with angular momentum 7/2,
providing sensitivity to all clock-frame parameters
$\tib_3^p$,
$\tic_q^p$,
$\tid_3^p$,
$\tig_d^p$,
$\tig_q^p$,
and so yielding both dipole and quadrupole shifts.
Among the components tested would be $c_{\mu\nu}^p$.
As a guide to what might be achieved,
we note that
an Earth-based experiment based on
the $\ket{4,4} \rightarrow \ket{4,3}$ transition
achieved the sensitivity level of
about 50 $\mu$Hz.\cite{lh}
A similar experiment on the ISS would be
reduced in duration by a factor of 16.
Furthermore,
an investigation of the double-frequency signal
$2\om_s$ would give access to
the spatial components of $c_{\mu\nu}^p$
at the $10^{-25}$ level,
and to other components at about the $10^{-21}$ level.
In all,
about 60 components of coefficients
for Lorentz and CPT violation would be accessible
with Planck-scale sensitivity.

\subsection{Rubidium Clocks}
Experiments with $^{87}$Rb
have similar features to experiments with $^{133}$Cs.
A suitable reference signal would be the
standard $\ket{2,0} \rightarrow \ket{1,0}$
clock transition,
which is insensitive to Lorentz and CPT violation,
while a Zeeman hyperfine transition such as
$\ket{2,1} \rightarrow \ket{2,0}$
could be used as a signal clock.
Due to its unpaired electron,
$^{87}$Rb has sensitivity to electron parameters
similar to that of an H maser or
a Zeeman hyperfine transition in $^{133}$Cs.
The sensitivity to proton parameters is
also analogous to that of $^{133}$Cs
up to factors of order unity,
because the Schmidt nucleon for $^{87}$Rb is a
proton with angular momentum 3/2.
The fact that the nuclear configuration has magic neutron number
means theoretical calculations may be more reliable
and that experimental results would be
cleaner.\cite{kla}
As with the case of $^{133}$Cs,
numerous Lorentz and CPT tests
sensitive to Planck-scale effects
could be done.

\subsection{Other Spacecraft}
Important Lorentz and CPT tests could also be done
with other types of spacecraft.
Of special interest would be missions
where the speeds of the craft with respect to the Sun are
larger than the $\be_s$ possible with satellites orbiting the Earth.
One example is the
planned
SpaceTime\cite{spacetime}
experiment,
which will attain $\be \simeq 10^{-3}$
on a solar-infall trajectory from Jupiter.
This mission
will fly co-located
$^{111}$Cd$^+$,
$^{199}$Hg$^+$,
and $^{171}$Yb$^+$ ion clocks
in a craft rotating several times per minute,
so that even 15 min might be long enough to
gather useful data for Lorentz and CPT tests.
For each of the three clocks,
the clock transitions
$\ket{1,0} \rightarrow \ket{0,0}$
are unaffected by Lorentz and CPT violation
and so could be used as reference signals.
A signal clock would run on
a Zeeman hyperfine transition such as
$\ket{1,1} \rightarrow \ket{1,0}$.
Sensitivity to electron parameters
would then be possible due to the electron
configuration.
All three clocks would have sensitivity to
the neutron parameters $\tib_3^n$, $\tid_3^n$, $\tig_d^n$
in the clock frame,
because the Schmidt nucleon for all three isotopes is
a neutron with angular momentum 1/2.
Such experiments are important
because none of the above neutron parameters can be probed with
the proposed ISS experiments.
By searching for variations in the signal clocks
at the spacecraft rotation frequency $\om_{ST}$
and also at $2\om_{ST}$
numerous tests
for Lorentz and CPT violation
would be possible.
Experiments of this type would have an order of magnitude
greater sensitivity to Lorentz and CPT violation
than measurements performed
either on the Earth
or in orbiting satellites
because of their large boost.

\section{Discussion}
There are numerous interesting prospects for investigating
CPT and Lorentz symmetry violation using space-based experiments.
These include experiments planned for the International Space Station
in the coming decade.
These experiments will be able to exploit the relatively high
rotation rates of the ISS as well as the relatively high
speed of motion around the Earth to gain sensitivity to
relativistic effects within the context of the standard-model extension.

\section*{Acknowledgments}
I thank collaborators Robert Bluhm, Alan Kosteleck\'y, and Chuck Lane.
This work was supported in part by a research grant
from Northern Michigan University.

\end{document}